----------------------------- Start of body part 4

\documentstyle[11pt]{article}

\begin{document}

\begin{titlepage}
\title{q - Magnetism at roots of unity}
\author{Alexander B\'erkovich and Germ\'an Sierra \\
Instituto de F\'{\i}sica Fundamental, CSIC \\
Serrano 123.  Madrid, SPAIN \\ \\
C\'esar G\'omez \\
D\'ept. Physique Th\'eorique, Universit\'e de Gen\`eve \\
CH-1211 Gen\`eve 4. Gen\`eve, SWITZERLAND}

\date{}
\maketitle

\begin{abstract}
We study the thermodynamic properties of a family of integrable
1D spin chain hamiltonians associated with quantum groups at roots
of unity. These hamiltonians depend for each primitive root of
unit on a parameter $s$ which plays the role of a continuous
spin. The model exhibits ferrimagnetism
even though the interaction involved is between nearest
neighbors. The latter phenomenon is interpreted as a genuine
quantum group effect with no ``classical" analog. The discussion of
conformal properties is given.
\end{abstract}

\vskip-14.0cm
\rightline{{\bf IMFF-6/92}}
\rightline{{\bf May 1992}}
\vskip2cm

\end{titlepage}

After Heisenberg [1], spin is the key word for understanding the magnetic
properties of metals. In 1 spatial dimension we have many exactly
solvable models, which can be treated by means of Bethe ansatz technique [2].
These models can be used to deepen our intuition on such non trivial subject
as magnetism. Quantum groups [3] provide the mathematical ground for
studying integrable one dimensional spin chains. Moreover, the different
integrable generalizations of the original $S=1/2$ Heisenberg model are
associated in one to one fashion with the different irreps of $U_q(SL(2))$
where the deformation
parameter $q$ is related to the anisotropy of the chain.

Heisenberg's ideas of magnetism can be extended naturally in the context of
quantum groups, in a sense that the rotational group $SU(2)$ is replaced by
$U_q(SL(2))$.
For generic $q$ this replacement is not essential, just because the finite
dimensional irreps of $SU(2)$ and the ones of its quantum deformation are
the same. If we want to get a typical signal of the effect of defining
the spin variables by finite dimensional irreps of $U_q(SL(2))$,
``q - magnetism",
we need to work in the very special regime, a $q$ root of unity, where we have
finite dimensional irreps of $U_q(SL(2))$ without ``classical" $(q=1)$
analog [4].

        In this letter we start a systematic study of the magnetic properties
of 1 dimensional spin chains using non regular finite dimensional irreps of
$U_q(SL(2))$ at roots of unity. The main new phenomena we find, concerning the
magnetic properties, is ferrimagnetism, i.e. a disordered ground state with
non vanishing magnetization. This kind of behavior is known in systems
possessing
complex topology of interaction [5], while here the appearance of this
phenomena
is directly tied with the special irreps used to define the spin variables of
the chain.

        The quantum group $U_q(SL(2))$ with $q=\epsilon$, $\epsilon^N=1$
is generated by the operators $E$, $F$ and $K=\epsilon^{2s_z}$. The peculiar
thing about $\epsilon$ being a root of unity is that $E^{N'}$, $F^{N'}$ and
$K^{N'}$
are central elements (where $N'=N$ if $N$ is odd and $N'=N/2$ if $N$ is even).
These central elements, together with the Casimir, label the irreps of
$U_q(SL(2))$. Regular irreps, which are the $q$-deformations of the usual
integer
and half-integers spin representations, satisfy $E^{N'}=F^{N'}=0$
and $K^{N'}=\pm 1$.
Nilpotent irreps of $U_{\epsilon}(SL(2))$ are a slight generalization of the
regular ones, in a sense, that the generator $K$ takes on the generic
value $\epsilon^{2s}$, where $s$ is our ``continuous" spin. The dimension of
these nilpotent irreps is always $N'$.

        The ``nilpotent" - spin chain hamiltonian is defined in the standard
way as:

\begin{equation}
 H(s)=\left. -iI\frac{\partial ln t(u,s)}
           {\partial u} \;\; \right|_{u=0}
\label{1}
\end{equation}

\noindent where $I$ is an overall coupling constant and $t(u,s)$ is the
transfer matrix defined by the quantum $R$-matrix $R^{ss}(u)$ intertwining
two nilpotent irreps of $U_{q}(\widehat{SL(2)})$ [6].

For the special case $N=3$ and $s=1$ the hamiltonian (1) coincides with
the Fateev-Zamolodchikov hamiltonian [7] with the anisotropy fixed by
$q=e^{2\pi i/3}$. The hermiticity condition on the hamiltonian $H(s)$ are
given by:

\begin{equation}
v_s \sin \gamma k \sin \gamma (2s-k+1)>0 \;\;\;\;\; k=1,2...,N'-1
\label{2}
\end{equation}

\noindent where $q=e^{i\gamma}$ and $v_s=\pm 1$ is the spin parity.
Equation (2) is equivalent to the condition $E^{\dagger}=v_sF$ for the
corresponding nilpotent $s$-irrep.
The hermiticity regions that follow from (2) (for $\epsilon=e^{2\pi i/N}$) are:

\begin{equation}
\begin{array}{rl}
N & even: \;\; \frac{1}{2}-\frac{1}{p_0}+\frac{1-v_s}{4} < \frac{s}{p_0} <
\frac{1}{2}+\frac{1-v_s}{4} \\
N & odd: \;\;\;\; \frac{1}{2}-\frac{3}{4p_0}+\frac{1-v_s}{4} < \frac{s}{p_0} <
\frac{1}{2}-\frac{1}{4p_0}+\frac{1-v_s}{4}
\end{array}
\label{3}
\end{equation}

\noindent where $p_0=\frac{N}{2}$. In what follows, we shall consider
$N>4$ ($N$ even) and $N>3$ ($N$ odd). In the trivial case $N=4$,
the hamiltonian
(1) is essentially that of $XX$-model in magnetic field. The case $N=3$ [8]
requires special treatment which will be given elsewhere [9].

        Notice that for $N$ even the middle point of both spin intervals
($v_s=\pm 1$) corresponds to a regular integer or half-integer spin. For $N$
odd only the interval of negative parity contains such a point. For all these
middle points $s_0$'s the corresponding hamiltonians $H(s_0)$ are identical
to a higher spin $XXZ$ models
with anisotropy $\gamma = \frac{2\pi}{N}$. It is interesting to observe that
$2s_0+1$ is not a Takahashi number [10].
Apparently for that reason, Kirillov and Reshetikhin
[11] do not consider this case in their, otherwise, general analysis.
On the other hand, Babujian and Tsvelick [12] have considered one of these
points ($s=\frac{N-2}{4}$ for $N$ even). However, we do not believe that their
results are correct at this point.

        The hamiltonian (1) can be diagonalized by means of the standard
Bethe ansatz [6]. The Bethe ansatz equations read in our case:

\begin{equation}
\left [
\frac{sh\frac{\gamma}{2}(\lambda_j+2is)}{sh\frac{\gamma}{2}(\lambda_j-2is)}
\right ]^L =-\prod_{k=1}^M \frac{sh\frac{\gamma}{2}(\lambda_j-\lambda_k+2i)}
{sh\frac{\gamma}{2}(\lambda_j-\lambda_k-2i)}
\label{4}
\end{equation}

\noindent with the energy eigenvalues given by:

\begin{equation}
E_M = - \sum_{k=1}^{M} \frac{I\sin 2\gamma s}{sh[\frac{\gamma}{2}
(\lambda_k+2is)]sh[\frac{\gamma}{2}(\lambda_k-2is)]}
\label{5}
\end{equation}

\noindent where $s$ is our ``generic" spin, subject only to hermiticity
requirements (3).

        To solve (4) we will use the String Hypothesis (SH) [13]:

\begin{equation}
\lambda_l^n = \lambda_c^n + i[n+1-2l+\frac{\pi}{2\gamma}(1-v_sv_n)]
\label{6}
\end{equation}

\noindent where $l=1,...,n$ and $\lambda_c^n$ is the real valued center
of the string of
length $n$ and parity $v_n=\pm 1$. It can be proven that the allowed
strings are determined by the Takahashi condition [10]:

\begin{equation}
v_n \sin \gamma (n-l) \sin \gamma l > 0 \;\;\;\;\; l=1,2...,n-1
\label{7}
\end{equation}

\noindent whenever the hermiticity condition (2) holds true.
Strictly speaking, SH is
legitimate only if the number of BA roots is much smaller than the number of
sites. However, it has been shown [14] that the SH can be safely used for the
nonzero magnetic field or temperature.

Using the ``Takahashi zone" terminology, we have for the
allowed strings $(n_j,v_j)$:

\begin{eqnarray}
N & {\rm even}: & \left\{ \begin{array}{lll}
0-{\rm zone} & n_j=j, v_{n_j}=+1 & 1\leq j \leq \nu-1 \\
1-{\rm zone} & n_{\nu}=1, v_{\nu}=-1 & j=\nu \\
\end{array}
\right. \nonumber \\
& & \label{8} \\
N & {\rm odd}: & \left\{ \begin{array}{ll}
0-{\rm zone} & \;\;\;\; n_j=j, v_j=+1
\;\;\;\;\;\;\;\;\;\;\;\;\;\;\;\;\;\;\;\;\;\; 1\leq j \leq \nu-1 \\
1-{\rm zone} & \left\{ \begin{array}{ll}
n_{\nu}=1, v_{\nu}=-1 & \;\;\; j=\nu \\
n_{\nu+1}=\nu+1, \; v_{\nu+1}=+1 & \;\;\; j=\nu+1 \\
\end{array}
\right. \nonumber \\
2-{\rm zone} & \;\;\;\; n_{\nu+2}=\nu, \;\;\; v_{\nu+2}=+1
\;\;\;\;\;\;\;\;\;\;\; j=\nu+2 \\
\end{array}
\right. \nonumber
\end{eqnarray}

\noindent where $\nu = \frac{N}{2}$ for $N$ even and $\nu=\frac{N-1}{2}$
for $N$ odd.

In the thermodynamic limit equations (4) become:

\begin{equation}
\tilde{a}_j = (-1)^{r_j} (\rho_j + \rho_j^h) + \sum_k T_{jk} \star \rho_k
\label{9}
\end{equation}

\noindent where $\rho_j(\rho_j^h)$ is the density of $j$-strings ($j$-holes)
and $(-1)^{r_j}$
is the sign of $\tilde{a}_j(\lambda)$ and $``\star"$ stands for convolution.
The Fourier transforms of the functions which appear in (9) are given by:

\begin{eqnarray}
\hat{T}_{jk} & = & g(\omega;|n_j-n_k|;v_{n_j}v_{n_k}) +
g(\omega;|n_j+n_k|,v_{n_j}v_{n_k})  \nonumber \\
&  & + 2 (1-\delta_{1,min(n_j,n_k)}) \sum_{l=1}^{min(n_j,n_k)-1}
g(\omega;|n_j-n_k|+2l;v_{n_j}v_{n_k}) \nonumber \\
\hat{\tilde{a_j}} & = & \sum_{l=0}^{n_j-1}g(\omega;2s+1-n_j+2l,v_sv_{n_j})
\label{10} \\
g(\omega;n;v) & = & - \frac{sh2p_0\omega ((\frac{n}{2p_0}+\frac{1-v}{4}))}
{shp_0\omega};
\;\;\;\;\;\;((x)) \;\;\; {\rm is\;\;Dedekind\;\;function} \nonumber
\end{eqnarray}

Following Yang and Yang [15] we minimize the free energy $F=E-TS$ to obtain:

\begin{eqnarray}
\frac{F}{L} & = & -T\sum_j \int_{-\infty}^{+\infty}
d\lambda|\tilde{a_j}(\lambda)|
ln(1+\eta_j^{-1}) \nonumber \\
\label{11} \\
ln\eta_j & = & -\frac{4p_0I}{T}\tilde{a_j}+\sum_k(-1)^{r_k} T_{jk}\star
ln(1+\eta_k^{-1}) \nonumber \\
\eta_j & = & e^{\frac{\epsilon_j(\lambda)}{T}}=
\frac{\rho_j^h(\lambda)}{\rho_j(\lambda)} \nonumber
\end{eqnarray}

In the $T=0$ limit we get the following results for the ground state and the
spectrum of excitations (see table 1).

\begin{center}
\begin{tabular}{|c|c|c|c|c|}
\hline
$N$ & $I$ & Ground state strings & Positive energy strings &
Zero energy strings \\
\hline
even & $>0$ & $\nu-1$ & $\nu$ & the rest \\
\hline
even & $<0$ & $\nu$ & the rest & none \\
\hline
odd & $>0$ & $\nu+2$ & $\nu,\nu+1$ & the rest \\
\hline
odd & $<0$ & $\nu,\nu+1$ & the rest & none \\
\hline
\end{tabular}
\end{center}
\begin{center}
Table 1.
\end{center}

\noindent The entries in the table above refer to the label $j$ of the strings
$(n_j,v_j)$.

We observe that this spectrum of the strings at zero temperature is independent
of the value of the spin $s$, as long as it belongs to the hermiticity
regions (3).
A comparison of the spectrum given above with that of ref. [11] shows that
they are quite different. Interestingly enough, there is only one
kind of string
filling the Dirac sea (except for the case of $N$ odd and $I<0$). This will be
important when we discuss the conformal properties of our models.

The $T\rightarrow\infty$ limit of equations (11) provides a justification
of the SH. In fact, we get $\lim_{T \to \infty}\frac{F}{TL}=-lnN'$,
which implies that the total number of states is correctly given by $(N')^L$.

Next we move on to compute entropy $S$:

\begin{equation}
\frac{S}{L}=\sum_j \int d\lambda\rho_j(\lambda)[(1+\eta_j)ln(1+\eta_j)-
\eta_jln\eta_j]
\label{12}
\end{equation}

\noindent Making use of equations (11), we obtain in low temperature limit:

\begin{eqnarray}
N \;\;\; {\rm even}: & \frac{S}{L} = & \left\{ \begin{array}{ll}
\frac{2T\pi}{6v^s}[3-\frac{6}{\nu+1}]; & I>0 \\
\frac{2T\pi}{6v^s}; & I<0 \\
\end{array}
\right. \nonumber \\
\label{13} \\
N \;\;\; {\rm odd}: & \frac{S}{L} = & \left\{ \begin{array}{ll}
\frac{2T\pi}{6v^s}[3-\frac{6}{\nu+2}]; & I>0 \\
\frac{4T}{\pi}[\frac{1}{v_1^s}L(\frac{\nu}{2\nu+1})+\frac{1}{v_2^s}
L(\frac{\nu+1}{2\nu+1})]; & I<0
\end{array}
\right. \nonumber
\end{eqnarray}

\noindent where $v^s,v_1^s$ and $v_2^s$ are speeds of sound:

\begin{eqnarray}
v^s=\frac{N}{2}|I|, & \;\;\;\; v_1^s=\frac{\frac{N}{2}|I|}{\frac{N}{2}-1}, &
\;\;\;\; v_2^s=N|I|
\label{14}
\end{eqnarray}

\noindent and $L(x)$ is the dilogarithmic Roger's function [16]. Notice that
for $N$ odd and $I<0$ we have two different speeds of sound. For the remaining
cases there is only one speed of sound so that the underlying CFT has a central
extension $c$ given by:

\begin{equation}
\frac{\partial S}{\partial T} \equiv -\frac{\partial^2F}{\partial T^2} =
\frac{\pi c}{3v^s}
\label{15}
\end{equation}

\noindent From equation (13) we get:
\begin{eqnarray}
\begin{array}{ll}
I>0, & \;\;\;\; c=\frac{3s_{eff}}{s_{eff}+1} \nonumber \\
I<0, & \;\;\;\; c=1 \;\;\;\;\;{\rm for\;\;}N\;\;{\rm even}
\end{array}
\label{16}
\end{eqnarray}

\noindent where

$$
s_{eff} = \left\{ \begin{array}{ll}
\frac{N-2}{4}, & \;\;\;\;N\;\;{\rm even} \\
\frac{N-1}{4}, & \;\;\;\;N\;\;{\rm odd}
\end{array}
\right. \nonumber
$$

When $N$ is odd and $I<0$ there are two different strings filling the
ground state and two different speeds of sound. This fact indicates that
rotational invariance is broken which, in turn, implies that we do not
have a full conformal invariance. This situation has already been
discussed in literature [17], where a broken CFT (in the sense given above)
can be viewed as a sum of two independent CFT's. In our case, we have not been
able to identify any of the broken pieces with reasonable CFT.

Finally, we present our results for the magnetization of the ground state at
$T=0$ which is defined as:

\begin{eqnarray}
M=\frac{s^z}{L}=s-\sum_{j\in\;\;{\rm Ground\;\;state}}
n_j \int \rho_j (\lambda) d \lambda
\label{17}
\end{eqnarray}

\noindent The results are collected in table 2:

\begin{center}
\begin{tabular}{|c|c|c|}
\hline
$N$ & $I$ & $M$ \\
\hline
even & $>0$ & $M=\frac{N}{2}[s-(\frac{N}{2}-1)\frac{3-v_s}{4}]$ \\
\hline
even & $<0$ & $M=\frac{N}{2}(\frac{1-v_s}{4})$ \\
\hline
odd & $>0$ & $M=N[s-(\frac{N-1}{2})\frac{3-v_s}{4}]$ \\
\hline
odd & $<0$ & $M=-N[s+1-(\frac{N+1}{2})\frac{3-v_s}{4}]$ \\
\hline
\end{tabular}
\end{center}
\begin{center}
Table 2.
\end{center}

\noindent From table 2, we see that for generic $s$ (subject to hermiticity
condition (3)) the ground state exhibits ferrimagnetic behavior.
More precisely,
when spin $s$ takes on values different from integer or half-integer then
magnetization is non null.

To summarize:  $q$ being a root of unity made it possible to depart from
regular representations and this, in turn, led to the new phenomenon of
ferrimagnetism for a system governed by local (nearest neighbors interaction)
hamiltonian. In our future publications we hope to report on our study of
magnetic properties of the model as well as on the further analysis of
conformal properties and to present our study of scattering matrices
along with quantum numbers of low lying excitations. The
details of the results presented here will be given elsewhere [9].

{\bf Acknowledgments.}

We are grateful to L. Nirenberg for the prompt typing of this manuscript.

\end{document}